\documentclass[twoside]{dis09}
\usepackage[latin1]{inputenc}
\usepackage[dvips]{graphicx,epsfig,color}
\usepackage{wrapfig,rotating}
\usepackage{amssymb,amsmath,array}

\newcommand{\mrf}[1]{\mbox{$\mathrm{#1}$}}
\newcommand{\mif}[1]{\mbox{$\mathit{#1}$}}

\newcommand{\DGG}{\mif{\Delta G/G}}

\newcommand{\gom}{\mrf{GeV/\mif{c}}}

\begin{document}
\title{COMPASS Results on Collins and Sivers Asymmetries}

\author{Andrea Bressan$^1$ \\ {\small (on behalf of the COMPASS collaboration)}
%
%
\vspace{.3cm}\\
%
1- Department of Physics, University of Trieste and Trieste Section of INFN,\\
via Valerio 2, 34127, Trieste, Italy
}

\maketitle

\begin{abstract}
In the list of the main items studied by the CERN COMPASS experiment there are
the transverse spin and momentum effects visible in the azimuthal distributions
of hadrons produced in the deep inelastic scattering.
In the years 2002-2004 COMPASS has collected data with a $^6$LiD target with the
polarization oriented transversely with respect to the muon beam direction for
about 20\% of the running time; in 2007, COMPASS has used for the first time a proton
NH$_3$ target with the data taking time equally shared between longitudinal and
transverse polarization of the target. After reviewing the results obtained with
the deuteron, the new results for the Collins and Sivers asymmetries of the
proton will be presented.
\end{abstract}

\section{Introduction}

The study of transverse spin and momentum effects of the nucleon have started
significantly later than the longitudinal case, mainly since there was a
prejudice that transverse spin is small or irrelevant for ultra-relativistic
particles, or, at least in hard reactions.  The large single-spin asymmetries in
pion production from the interaction of a transversely polarized 200 GeV protons
on a liquid \mrf{H_2} target first reported by the Fermilab E704
collaboration~\cite{e704} (and today confirmed at higher energies by the RHIC
experiments) were not understood but believed to be a tail of low-energy
phenomena.  During the nineties both the experimental and the theoretical
progress allowed to better spot the relevance of transverse spin and momentum
effects for a deeper understanding of the nucleon structure.  As a consequence,
at that time several new experiments (HERMES at DESY, COMPASS at CERN, the RHIC
experiments at BNL) were proposed; since that, the activity in the field is
continuously growing, both theoretically and experimentally, giving a more and
more complete picture.  It is now well established that to fully specify the
quark structure of the nucleon at the twist-two level, the transversity
distributions $\Delta_T q(x)$ have to be added to the unpolarised distributions
$q(x)$ and the helicity distributions $\Delta q(x)$. The transversity PDF's give
the probability that the quark spins are aligned parallel or antiparallel to the
spin of a transversely polarised nucleon. They are difficult to measure, since
they are chirally odd and need to be coupled to a chirally odd partner. In
particular, they cannot be measured in inclusive deep-inelastic scattering
(DIS). In hadron colliders (or with hadronic beams) they can be measured looking
at the Drell-Yan processes as proposed by the RHIC experiments~\cite{rhic1} for
transversely polarized protons on proton scattering and by GSI
experiments~\cite{gsi1} with hard polarised proton anti-proton scattering.  They
can also be measured in semi-inclusive DIS (SIDIS) of leptons on transversely
polarised nucleons in which final state hadrons are also detected.  To access
the transversity PDF in SIDIS, one has to measure the quark polarisation, i.e.
to use the so-called 'quark polarimetry'. Different techniques have been
proposed in so far.  Three of them are presently used in COMPASS, namely:
\begin{itemize}
\item[-] measurement of the single-spin asymmetries (SSA) in the azimuthal
distribution of the final state hadrons (the so-called Collins asymmetry);
\item[-] measurement of the polarisation of final state hyperons (the so-called
$\Lambda$ polarimetry);
\item[-] measurement of the SSA in the azimuthal distribution of the plane
containing the final state hadron pairs (the so-called two-hadron asymmetry);
\end{itemize}
In this contribution the results obtained by COMPASS with the first two methods
will be given while the results obtained with the last of the three polarimeters
is presented in~\cite{heiner}.

The Collins asymmetry \mrf{A_{Coll}}, is due
to the combined effect of $\Delta_T q$ and the chiral-odd Collins fragmentation
function $\Delta^0_T D^h_q$, which describes the spin-dependent part of the
hadronization of a transversely polarized quark into a hadron with transverse
momentum $p_T^h$. At leading order, the Collins mechanism~\cite{Collins:1992kk}
leads to a modulation in the azimuthal distribution of the produced hadrons
given by:
\[
N(\Phi_{C}) = \alpha \cdot N_0 \, (1 + A\mrf{_{Coll}} \cdot P_T
\cdot f \cdot D_{NN} \sin{\Phi_{C}})\, ,
\]
\noindent where $\alpha$ contains the apparatus efficiency and acceptance, $P_T$
is the target polarization, $D_{NN}$ is the spin transfer coefficient and $f$ is
the fraction of polarizable nucleons in the target; $\Phi_C =\phi_h-\phi_{S'} =
\phi_h+\phi_S - \pi$ is the Collins angle, with $\phi_h$ the hadron azimuthal
angle, $\phi_{S'}$ the final azimuthal angle of the quark spin and $\phi_S$
the azimuthal angle of the nucleon spin in the $\gamma-N$ system. Finally
\[
A\mrf{_{Coll}} = \frac{\sum_q e^2_q \cdot \Delta_T q(x) \cdot  \Delta^0_T
  D_q^h(z,p_T^h)}{\sum_q e^2_q \cdot q(x) \cdot  D^h_q(z,p_T^h)}  
\]
\noindent is the Collins asymmetry.

Another way to access transversity is by measuring the transverse $\Lambda$ and
$\bar{\Lambda}$ polarization in the reaction $\mu N^\uparrow \rightarrow \mu
\Lambda^\uparrow X$. If the struck quark fragments into a $\Lambda$
hyperon in this reaction, the corresponding polarization is given by:
\[
P_\Lambda = f P_T D_{NN} (y) \frac{\sum_q e^2_q \cdot \Delta_T q(x) \cdot  \Delta_T
  D_q^\Lambda(z)}{\sum_q e^2_q \cdot q(x) \cdot  D^\Lambda_q(z)}
\]
\noindent where the T-axis for the measurement of the polarization is given by the
polarization vector of the struck quark.
In this case the transversity distributions appear coupled to the
chiral-odd part of the $\Lambda$ fragmentation functions $\Delta_T
D^\Lambda_q(z)$, which are so far completely unknown.  

Another important aspect under study in this field is the role of the quark
intrinsic transverse momentum and the connection with the spin for the
description of the nucleon structure. The transverse momentum dependent (TMD)
distribution functions (PDF) and fragmentation functions (FF) are today
considered an important ingredient in the structure of the nucleon. The SIDIS
cross-section in one-photon exchange approximation contains eight TMD PDF, three
of which survive upon integration over the transverse momenta.  Some of these
TMD distributions can be extracted in SIDIS looking at the azimuthal
distributions of the final state hadrons. This is particularly true for the
so-called Sivers asymmetry, which is, together with the Collins asymmetry,
presently the most studied. Through the Sivers asymmetry it is possible to
access the Sivers PDF, which takes into account a possible deformation in the
distribution of the quark intrinsic transverse momentum in a transversely
polarised nucleon. Particularly interesting is also the so called Boer-Mulders
function and the COMPASS efforts to access it are described
in~\cite{bressanCAHN}. Measuring SIDIS on a transversely polarized target allows
the Collins and the Sivers effects to be disentangled.  
The Sivers asymmetry can be written as:
\[
A\mrf{_{Siv}} = \frac{\sum_q e^2_q \cdot \Delta^T_0 q(x,p_T^h) \cdot 
  D^h_q(z)}{\sum_q e^2_q \cdot q(x) \cdot  D^h_q(z)} 
\]
with a modulation expressed in terms of the Sivers angle $\Phi_S =
\phi_h-\phi_S$. Since in this case the unpolarized fragmentation functions are
known, the measurement of the Sivers asymmetry for both positive and negative
produced hadrons allows a direct extraction of the Sivers functions, if the
measured asymmetry are different from zero, while a zero result for an isoscalar
target like the $^6$LiD used in COMPASS can come both from a vanishing Sivers
function or from a cancellation between $u$ and $d$ quark contributions.

\section{The COMPASS experiment} 

The COMPASS experiment has been set up at the CERN SPS M2 beam line.  It
combines high rate beams with a modern two stage magnetic
spectrometer\cite{nimcompass}. \\ 
\begin{wrapfigure}{r}{0.60\columnwidth}
\vspace{-20pt}
\begin{center}
\includegraphics[width=0.59\columnwidth]{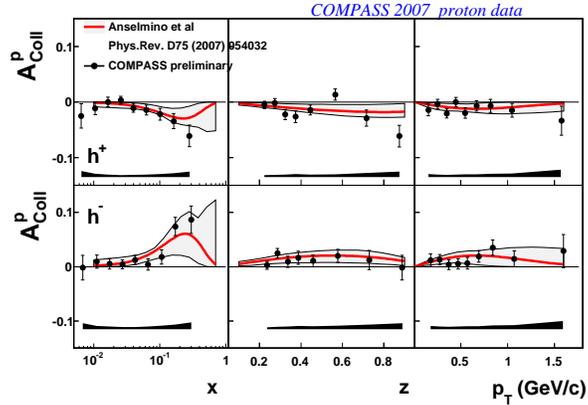}
\end{center}
\vspace{-20pt}
\caption{Collins asymmetry for positive (upper row) and negative
  (lower row) hadrons as a function of $x$, $z$ and $p_T$ for the proton 2007 data.}
\label{resultsColl}
\end{wrapfigure}
Both stages are equipped with tracking devices covering the full acceptance,
with e.m. and hadronic
calorimetry and muon identification via filtering through thick absorbers. In
the first stage a RICH detector is also installed, allowing the identification
of charged hadrons up to 50 GeV.  Detectors, electronics and data acquisition
system are able to handle beam rates up to 10$^8$ muons/s.
 The triggering system and the tracking system of
COMPASS have been designed to stand the associated rate of secondaries, and use
state-of-the-art detectors.  Also, fast front-end electronics, multi-buffering,
and a large and fast storage of events are essential.

COMPASS has collected data with a 160 GeV positive
muon beam impinging on a polarized solid target. The beam is naturally polarized by the
$\pi -$decay mechanism, and the beam polarization is estimated to be $\sim 80\%$
with a $\pm 5\%$ relative error.  The beam intensity is $2\times 10^8$ muons per spill.

Up to 2004 COMPASS has used the polarized target system of the SMC experiment,
which allows for two oppositely polarized target cells, 60~cm long each.  The PT
magnet can provide both a solenoid field (2.5 T), to hold the longitudinal (with
respect to the beam direction) polarization, and a dipole field (0.5 T), needed
for adiabatic spin rotation and for holding the transverse polarization.  In
2006 the installation of a new PT magnet, with an increased inner bore radius
matching the full acceptance of the spectrometer (180 mrad), was performed.

Up to 2006 the experiment has used $^6$LiD as deuteron target because its
favorable dilution factor of $\simeq$0.4, particularly important for the
measurement of \DGG. In 2007 an ammonia NH$_3$ target has been used as proton
target. Moreover, the target material has been distributed in three cells, with a
length of 30 cm for the outer cells and 60 cm for the inner one. In this case
the outer cells have the same orientation of the polarization, opposite to the
central one. Polarizations of 50\% and 90\% have been reached, respectively for the
two target materials.

\section{Analysis and Results}
\begin{wrapfigure}{r}{0.60\columnwidth}
\vspace{-20pt}
\begin{center}
\includegraphics[width=0.59\columnwidth]{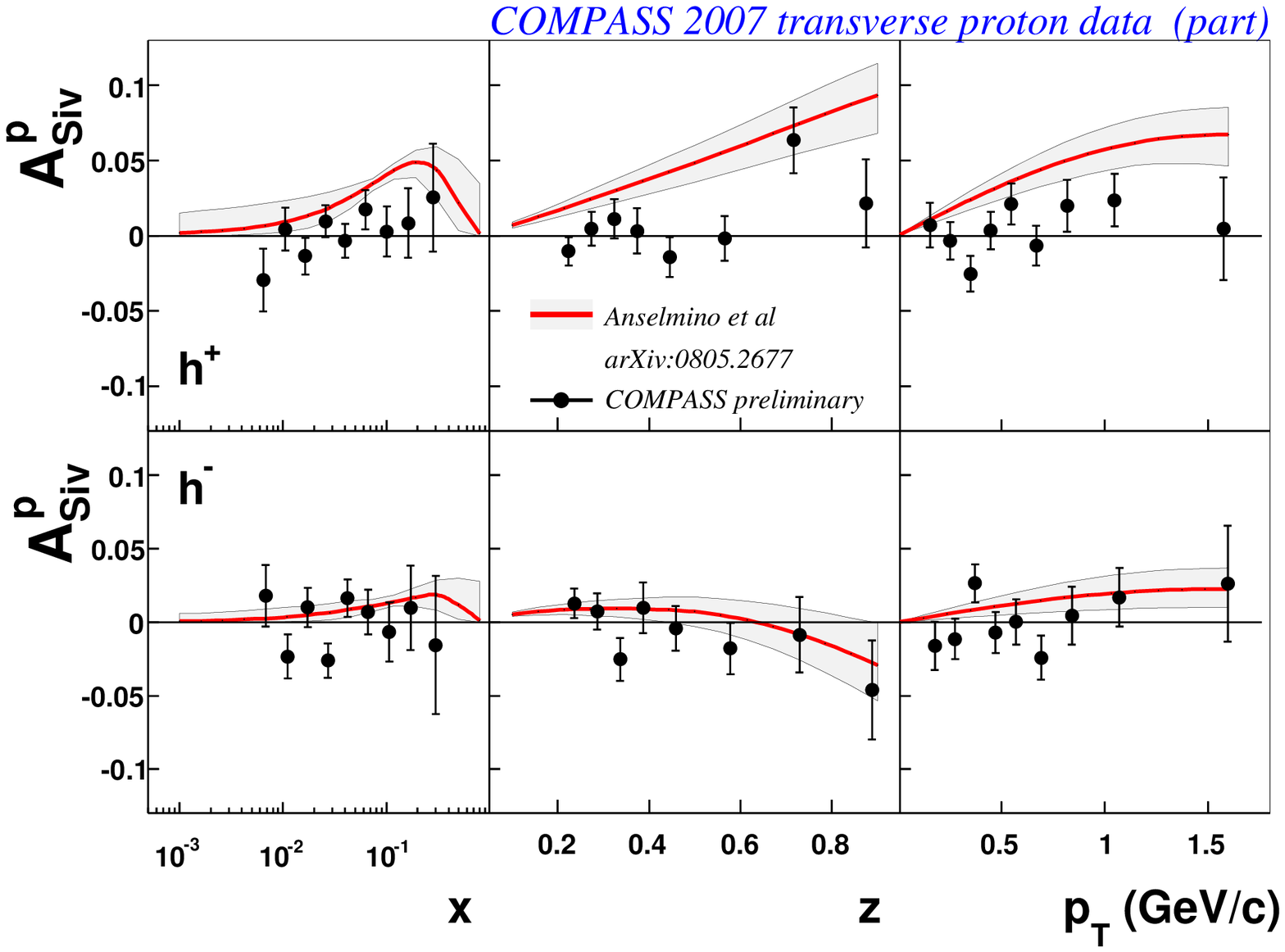}
\end{center}
\vspace{-20pt}
\caption{Sivers asymmetry for positive (upper row) and negative
  (lower row) hadrons as a function of $x$, $z$ and $p_T$ for the proton 2007 data.}
\label{resultsSiv}
\end{wrapfigure}
The event selection requires standard DIS cuts, i.e. $Q^2 > 1\ (\gom)^2$,
mass of the final hadronic state $W>5\ \mrf{GeV}/c^2$, $0.1 < y < 0.9$, and the
detection of at least one hadron in the final state.  For the detected hadrons
it is also required that: 
\begin{itemize}
\item the fraction of the virtual photon energy carried is
$z=E_h/E_\gamma>0.2$ to select hadrons from the current fragmentation region;
\item
$p_T > 0.1\ \mrf{GeV}/c$ (where $p_T$ is the hadron transverse momentum with
respect to the virtual photon direction) for a better determination of the
azimuthal angle $\phi_h$. 
\end{itemize}
\begin{wrapfigure}{r}{0.60\columnwidth}
\vspace{-20pt}
\begin{center}
\includegraphics[width=0.59\columnwidth]{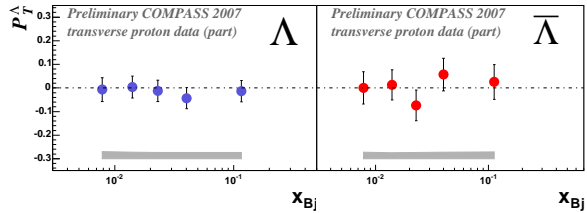}
\end{center}
\vspace{-20pt}
\caption{Transverse $\Lambda$ and $\bar{\Lambda}$ polarization as a function of
  $x$ for part of the 2007 proton data.}
\label{resultsLambda}
\end{wrapfigure}
The asymmetries have been calculated as a function of $x$, $z$ and $p_T$ for
positive and negative hadrons respectively.  Both the resulting Collins and
Sivers asymmetries from the whole deuteron data turned out to be small and
compatible with zero~\cite{npb} (a trend that is also shown by the identified
hadron results~\cite{plbid}), a result which was interpreted as a cancellation
between the contribution of the $u$ and $d$ quarks, for the isoscalar deuteron
target. The new results for the proton \mrf{NH_3} target are shown in
Fig.~\ref{resultsColl} for the Collins asymmetries and in Fig.~\ref{resultsSiv}
for the Sivers asymmetries, together with the prediction
from~\cite{anselminoCol,anselminoSiv}, based on the global analysis of the
HERMES proton data, COMPASS deuteron data and BELLE $e^+e^-$ data for Collins
and on the HERMES proton data and COMPASS deuteron data for Sivers. 
Collins asymmetries as a function of $x$ are small, compatible with zero, up to
$x~\sim0.05$, while in the last points a signal appears, and the asymmetries
increas up to 10\% with opposite sign for the positive (upper row) and
negative (lower row). The trend is in good agreement with what observed by  
HERMES~\cite{hermes}. At variance the Sivers asymmetries are small and
compatible with zero over the full $x$ range and for both positive and negative
hadrons; in this case the compatibility with HERMES results is fine for negative
hadrons but is marginal, if any, for positive hadrons. The origin of the
disagreement, needs to be understood~\cite{vogelsang} and will be an interesting
issue for the near future.

In Fig.~\ref{resultsLambda} the measured transverse $\Lambda$ and $\bar{\Lambda}$
polarizations are shown as a function of $x$. The result is statistically
compatible with zero, over the full range, and this may come both from the
relatively low-$x$ region sampled, or by the fact that most of the statistics
comes from a region with $z_\Lambda < 0.4$, where $\Delta_T D^\Lambda_q$ may be
small.  




\begin{footnotesize}

\end{footnotesize}


\end{document}